\documentclass[aps,prl,preprint,tightenlines,superscriptaddress,showpacs,byrevtex]{revtex4}
\usepackage{graphicx} 
\usepackage{dcolumn}  
\usepackage{rotating}
\usepackage{color}

\graphicspath{{ps}}

\begin{document}


\preprint{\vbox{ \hbox{   }
                 \hbox{  }
                 \hbox{BELLE-CONF-0558}
                 \hbox{EPS05-533} 
}}

\title{ \quad\\[0.5cm]  Moments of the Electron Energy Spectrum in $B \to X_c \ell \nu$ decays at Belle }

%
\affiliation{Aomori University, Aomori}
\affiliation{Budker Institute of Nuclear Physics, Novosibirsk}
\affiliation{Chiba University, Chiba}
\affiliation{Chonnam National University, Kwangju}
\affiliation{University of Cincinnati, Cincinnati, Ohio 45221}
\affiliation{University of Frankfurt, Frankfurt}
\affiliation{Gyeongsang National University, Chinju}
\affiliation{University of Hawaii, Honolulu, Hawaii 96822}
\affiliation{High Energy Accelerator Research Organization (KEK), Tsukuba}
\affiliation{Hiroshima Institute of Technology, Hiroshima}
\affiliation{Institute of High Energy Physics, Chinese Academy of Sciences, Beijing}
\affiliation{Institute of High Energy Physics, Vienna}
\affiliation{Institute for Theoretical and Experimental Physics, Moscow}
\affiliation{J. Stefan Institute, Ljubljana}
\affiliation{Kanagawa University, Yokohama}
\affiliation{Korea University, Seoul}
\affiliation{Kyoto University, Kyoto}
\affiliation{Kyungpook National University, Taegu}
\affiliation{Swiss Federal Institute of Technology of Lausanne, EPFL, Lausanne}
\affiliation{University of Ljubljana, Ljubljana}
\affiliation{University of Maribor, Maribor}
\affiliation{University of Melbourne, Victoria}
\affiliation{Nagoya University, Nagoya}
\affiliation{Nara Women's University, Nara}
\affiliation{National Central University, Chung-li}
\affiliation{National Kaohsiung Normal University, Kaohsiung}
\affiliation{National United University, Miao Li}
\affiliation{Department of Physics, National Taiwan University, Taipei}
\affiliation{H. Niewodniczanski Institute of Nuclear Physics, Krakow}
\affiliation{Nippon Dental University, Niigata}
\affiliation{Niigata University, Niigata}
\affiliation{Nova Gorica Polytechnic, Nova Gorica}
\affiliation{Osaka City University, Osaka}
\affiliation{Osaka University, Osaka}
\affiliation{Panjab University, Chandigarh}
\affiliation{Peking University, Beijing}
\affiliation{Princeton University, Princeton, New Jersey 08544}
\affiliation{RIKEN BNL Research Center, Upton, New York 11973}
\affiliation{Saga University, Saga}
\affiliation{University of Science and Technology of China, Hefei}
\affiliation{Seoul National University, Seoul}
\affiliation{Shinshu University, Nagano}
\affiliation{Sungkyunkwan University, Suwon}
\affiliation{University of Sydney, Sydney NSW}
\affiliation{Tata Institute of Fundamental Research, Bombay}
\affiliation{Toho University, Funabashi}
\affiliation{Tohoku Gakuin University, Tagajo}
\affiliation{Tohoku University, Sendai}
\affiliation{Department of Physics, University of Tokyo, Tokyo}
\affiliation{Tokyo Institute of Technology, Tokyo}
\affiliation{Tokyo Metropolitan University, Tokyo}
\affiliation{Tokyo University of Agriculture and Technology, Tokyo}
\affiliation{Toyama National College of Maritime Technology, Toyama}
\affiliation{University of Tsukuba, Tsukuba}
\affiliation{Utkal University, Bhubaneswer}
\affiliation{Virginia Polytechnic Institute and State University, Blacksburg, Virginia 24061}
\affiliation{Yonsei University, Seoul}
  \author{K.~Abe}\affiliation{High Energy Accelerator Research Organization (KEK), Tsukuba} 
  \author{K.~Abe}\affiliation{Tohoku Gakuin University, Tagajo} 
  \author{I.~Adachi}\affiliation{High Energy Accelerator Research Organization (KEK), Tsukuba} 
  \author{H.~Aihara}\affiliation{Department of Physics, University of Tokyo, Tokyo} 
  \author{K.~Aoki}\affiliation{Nagoya University, Nagoya} 
  \author{K.~Arinstein}\affiliation{Budker Institute of Nuclear Physics, Novosibirsk} 
  \author{Y.~Asano}\affiliation{University of Tsukuba, Tsukuba} 
  \author{T.~Aso}\affiliation{Toyama National College of Maritime Technology, Toyama} 
  \author{V.~Aulchenko}\affiliation{Budker Institute of Nuclear Physics, Novosibirsk} 
  \author{T.~Aushev}\affiliation{Institute for Theoretical and Experimental Physics, Moscow} 
% ...
\collaboration{The Belle Collaboration}
 

\noaffiliation

\begin{abstract}
We report a measurement of the inclusive electron energy spectrum for charmed semileptonic decays of $B$ mesons in a  $140\,{\rm fb}^{-1}$ data sample collected on the $\Upsilon(4S)$ resonance,
with the Belle detector at the KEKB asymmetric energy $e^+ e^-$ collider.  We determine the first, second and third moments of the electron energy spectrum for threshold values of the electron energy between 0.4 and 1.5 GeV.
\end{abstract}


\maketitle

\tighten

{\renewcommand{\thefootnote}{\fnsymbol{footnote}}}
\setcounter{footnote}{0}

\section{Introduction}

The Cabibbo-Kobayashi-Maskawa matrix element $|V_{cb}|$ can be extracted from the inclusive
branching fraction for charmed semileptonic $B$-meson decays $\mathcal{B}( B \to X_c \ell \nu )$\cite{bigi,wise}.  Several studies have shown that the spectator model decay rate is the leading term in a well-defined expansion controlled by  the parameter $\Lambda _{\rm QCD}/m_b$. Non-perturbative corrections to this leading approximation arise only at order $1/m_b^2$ and above. The key issue in this approach is the ability to separate non-perturbative corrections, that can be expressed as a series in powers of $1/m_b$, and perturbative corrections, expressed in powers of 
$\alpha _s$.

The coefficients of the $1/m_b$ power terms are expectation values of operators that include non-perturbative physics. Different expansions exist, reflecting a difference in the
approach used to handle the energy scale $\mu$ which separates
long-distance from short-distance physics~\cite{ref:1}. 

The shape of the lepton spectrum provides constraints on the heavy quark expansion based on local Operator Product Expansion (OPE) \cite{ref:2}, which calculates properties of the $B \to X_c \ell \nu$ transitions.  So far, measurements of the electron energy distribution have been made by DELPHI, CLEO, BABAR and Belle collaborations \cite{ref:4,cleoel,babarel,me1}.  In this note we report a measurement of the first,  second and third moment of the electron energy spectrum with a minimum electron momentum cut ranging between 0.4 and 1.5 GeV/$c$ in the $B$ meson rest frame.  

The data used in this analysis were collected with the Belle detector
at the KEKB~\cite{KEKB} asymmetric energy $e^+ e^-$ collider.
The Belle~\cite{Belle} detector is a large-solid-angle magnetic spectrometer that
consists of a three-layer silicon vertex detector (SVD), a 50-layer central drift chamber (CDC), 
an array of aerogel threshold \v{C}erenkov counters (ACC), 
a barrel-like arrangement of time-of-flight scintillation counters (TOF), and an electromagnetic calorimeter comprised of CsI(Tl) crystals (ECL) located inside a super-conducting solenoid coil that provides a 1.5~T magnetic field.  An iron flux-return located outside of
the coil is instrumented to detect $K_L^0$ mesons and to identify muons (KLM).

The present results are based on a $140\,{\rm fb}^{-1}$ data sample collected at the $\Upsilon (4S)$ resonance (on-resonance), which contains $1.5 \times 10^8$ $B \bar B$ pairs.  An additional $15\,{\rm fb}^{-1}$ data sample taken at 60MeV below the $\Upsilon (4S)$ resonance (off-resonance) is used to perform background subtraction from the $e^+e^- \rightarrow q \bar q$ process.  Events are selected by  fully reconstructing one of the $B$ mesons, produced in pairs from $\Upsilon (4S)$ decays.

We used a fully simulated generic Monte Carlo sample equivalent to 2.4 times the on-resonance integrated luminosity.  Simulated Monte Carlo events are generated with the EVTGEN event generator, and full detector simulation based on GEANT is applied~\cite{BelleMC}.  

\section{Event Selection}
After selecting hadronic events~\cite{hadron}, we fully reconstruct the decay of a $B$ meson on one side (tag-side), in
the decay modes $B \to D^{(*)} \pi^+, D^{(*)} \rho^+, D^{(*)} a_1^+$,
yielding a high purity $B$ meson sample. The following sub-decay modes are
reconstructed:

\begin{itemize}
\item $\bar{D}^{*0}\to \bar{D}^0\pi^0, \bar{D}^0\gamma$,
\item $D^{*-}\to \bar{D}^0\pi^+, D^-\pi^0$,
\item $\bar{D}^0\to K\pi, K\pi\pi^0, K\pi\pi\pi, K_S\pi\pi, K_S\pi^0$ and
\item $D^-\to K\pi\pi, K_S\pi$.
\end{itemize}

For each selected event, we calculate the beam-constrained mass $M_{\mathrm{bc}}$ and energy difference $\Delta E$:
\begin{equation}
  M_{\mathrm{bc}} = \sqrt{(E^*_{\mathrm{beam}})^2-(p^*_B)^2}, \quad \Delta E =
  E^*_B-E^*_{\mathrm{beam}},
\end{equation}
where $E^*_{\mathrm{beam}}$, $p^*_B$ and $E^*_B$ are the beam energy, the
$B$ momentum and the $B$ energy in the centre of mass
frame, respectively. 
Events with   $M_{\mathrm{bc}}>5.27~\mathrm{GeV/}c^2$  and $-0.06~\mathrm{GeV}<\Delta E< 0.08~\mathrm{GeV}$ are considered to be signal candidates.
The number of events, after continuum subtraction, in the  signal region are $63155\pm931$ and $40032\pm475$, $B^+$  and $B^0$  candidates, respectively \cite{CC}.  

\section{Electron Selection}

We search for electrons produced in semileptonic $B$ decays on the ``non-tag'' side. 
Particle tracks are selected if they originate from near the interaction vertex.
In addition, we measure tracks which pass through
 the barrel region of the detector, corresponding to an
angular acceptance of $35^{\circ} \leq \theta_{\mathrm{lab}} \leq
125^{\circ}$, where $\theta_{\mathrm{lab}}$ is the polar angle of the
track relative to the $z$ axis (opposite to the positron beam line).

Tracks that pass the above selection criteria and are not used in the reconstruction of the tag-side $B$ meson are considered as electron candidates. Electron identification is based on a
combination of ionisation $dE/dx$ measurements in the CDC, the response of the ACC, the shower shape in the ECL, and the ratio of energy deposited in the ECL to the momentum measured by the
tracking system ($E/p$) \cite{Belle}. 
The electron momentum spectrum is corrected by a momentum dependent electron detection efficiency, which includes the detector acceptance as well as the selection efficiency.
The momentum of the selected electrons is calculated in the $B$ meson rest frame ($p^{*B}_{e}$), exploiting the knowledge of the momentum of the fully reconstructed $B$. We require $p^{*B}_{e} \geq 0.4$~GeV/$c$.
The electron yield after these cuts is given in Table \ref{bzero}, with a signal purity of 69.6\% (50.0\%) for $B^+$ ($B^0$) tags.

Electrons suffer a loss of energy due to bremsstrahlung radiation in
the detector material in front of the calorimeter. This biases the
electron energy distribution. Hence, we partially correct for this
energy loss by recovering some of the emitted photons to restore the
original electron energy. The photon candidate is combined with the electron if the photon energy is below 1 GeV and the angle between the photon and the electron is less than 0.05 radians. 
  

 In  $B^+$ decays, prompt semileptonic decays ($b \rightarrow x \ell \nu$) of the non-tag side $B$ mesons are separated from cascade charm decays ($b \rightarrow c \rightarrow y \ell \nu$), based on the correlation between the flavor of the tagging $B$ and the electron charge. In $B^0$ decays, part of the sample mixes, which flips the correlation to the $B^0$. Thus in the $B^0$ sample we do not cut on the electron charge $-$ $B$ flavor correlation.

\section{Background Subtraction}

The reconstructed electron energy spectrum is contaminated by background processes, which should be evaluated and subtracted from the distribution before the extraction of the moments. The residual background is from:
\begin{itemize}
\item continuum background,
\item combinatorial background,
\item background from secondary decays,
\item $J/ \psi, \psi (2S)$, Dalitz decays and photon conversions, 
\item fake electrons.
\end{itemize}

The shape of the continuum background is derived from off-resonance data, and is normalised using the off- to on-resonance luminosity ratio and cross section difference. 
To account for low statistics in the off-resonance data we fit an exponential to the distribution, before rescaling.  The shape of the combinatorial background is derived from generic $B \bar B$ Monte Carlo events where either reconstruction or flavor assignment of the tagged $B$ meson is not carried out correctly.  The yield of this background is normalised to the on resonance data $M_{\mathrm{bc}}$ side band ($M_{\mathrm{bc}}<5.25~\mathrm{GeV/}c^2$), after the continuum background subtraction.
We also correct for cases where the fully reconstructed $B$ is correctly tagged, but the electron candidate either is not from a $B$ decay (secondary) or is a mis-identified hadron. These
background sources are irreducible; to estimate their magnitude we normalise the $B \bar B$ Monte Carlo using a fit to the electron momentum distribution after continuum and combinatorial background subtraction.  

\begin{figure}[htb!]
  \begin{tabular}{cc}
    \includegraphics[width=0.47\textwidth]{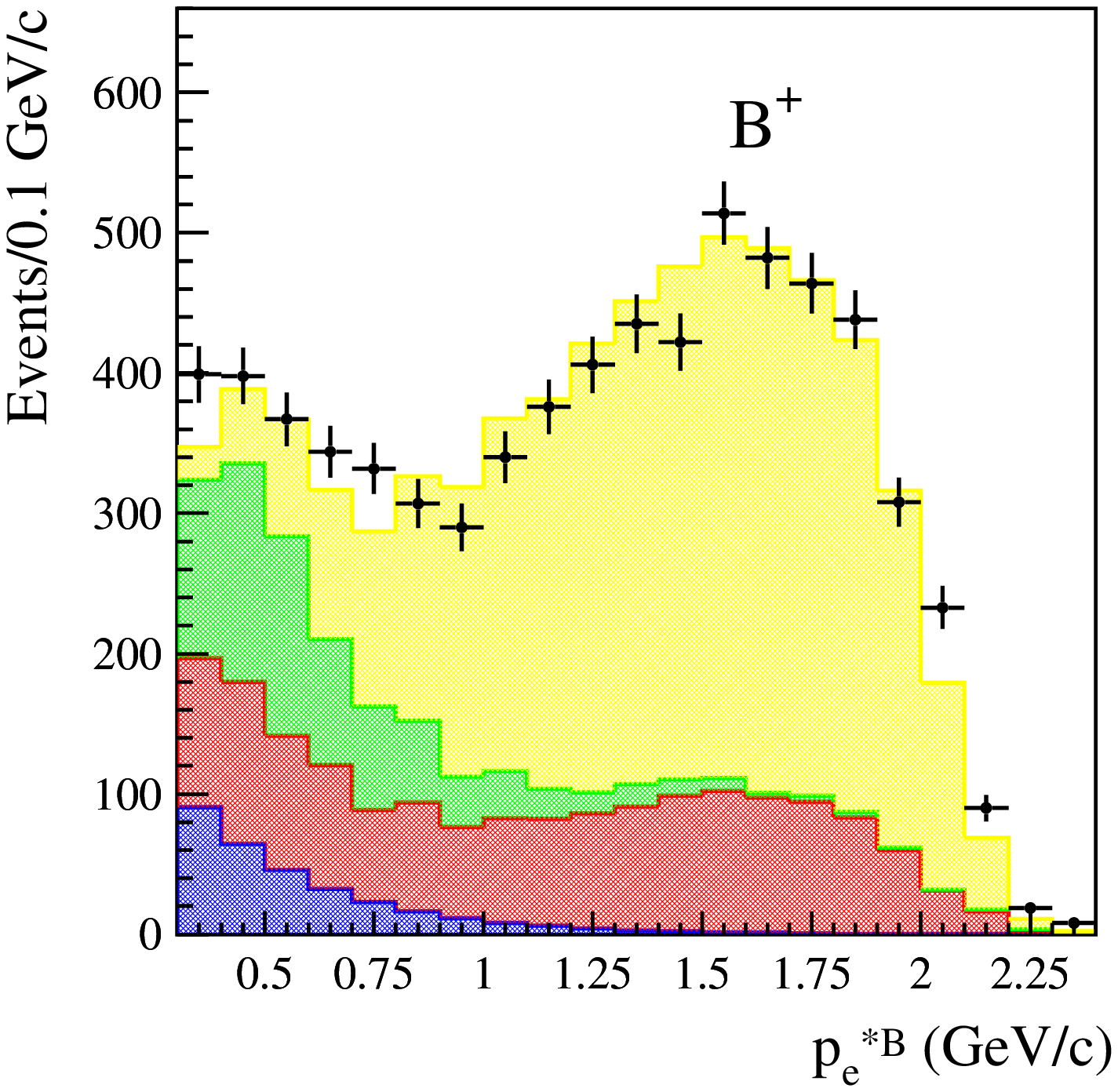}
    \includegraphics[width=0.47\textwidth]{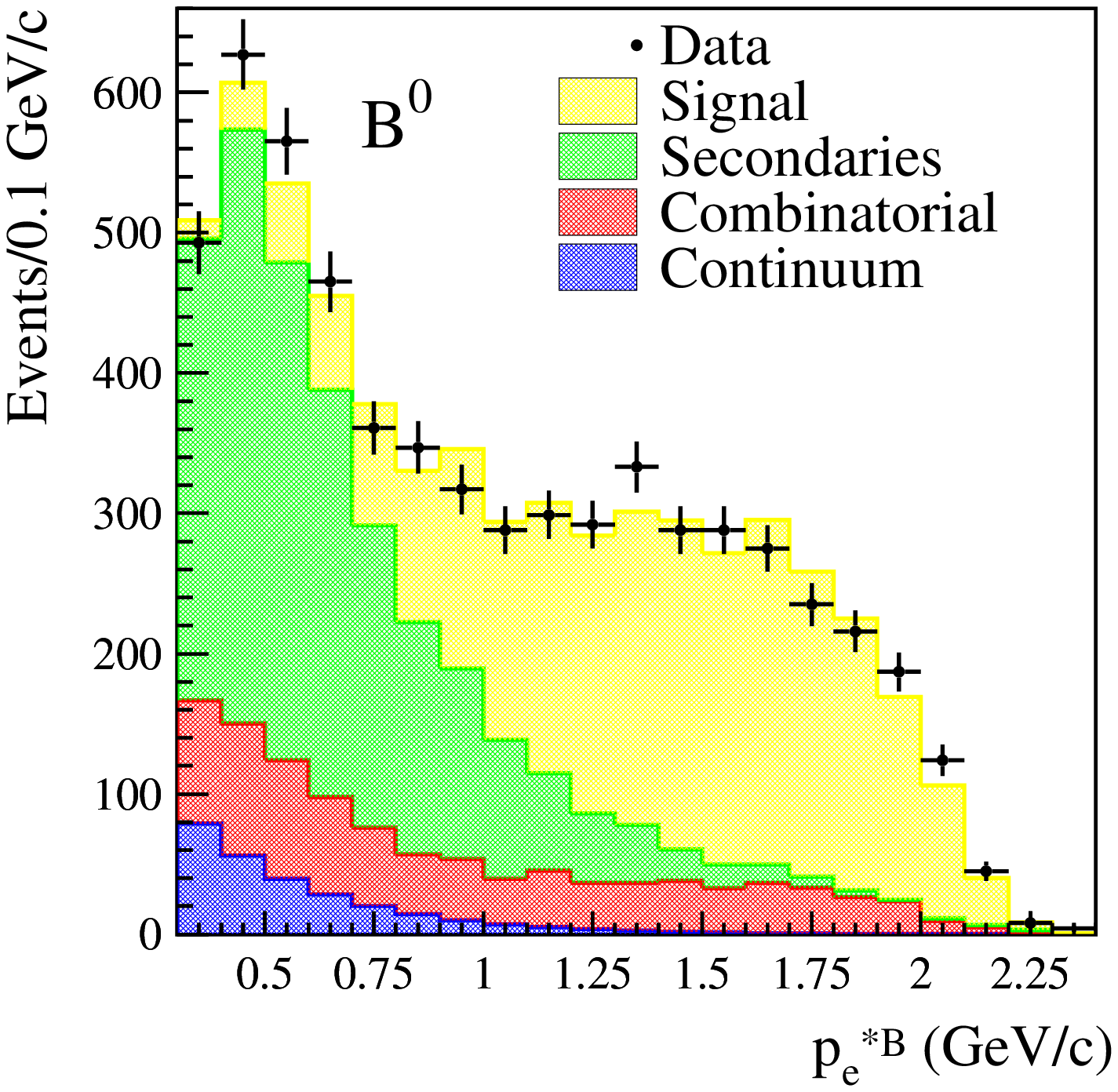} & \\
\end{tabular}
\caption{Breakdown of the backgrounds in the electron momentum spectra
  for $B^+$ , and $B^0$ 
  electrons. The Monte Carlo sample does not include $b \to u$ transitions. The errors shown are statistical only. }
\label{rawwithbg}
\end{figure}

The background distribution from $B \to D^{(*)} \to e$  decays are scaled using the latest published branching fractions~\cite{PDG}.  Contributions from $J/ \psi, \psi (2S)$ decays, photon conversions, and Dalitz decays are small after the track and electron selection cuts.  The surviving backgrounds are estimated using  Monte Carlo
simulation and subtracted along with the major secondary backgrounds.
The normalisation for the Monte Carlo yield of secondary and fake
leptons is obtained from data by fitting to the lepton momentum distribution in the range 0.3 GeV/$c$ $ < p_e^{*B} < $ 2.4 GeV/$c$.  Figure \ref{rawwithbg} shows the raw electron momentum spectrum with
all background contributions overlaid.  Table \ref{bzero} summarises the number of detected electrons and the contributions from these backgrounds.

\begin{table}[htb]
\caption{Electron yields for $p^{*B}_{e} \geq 0.4$ GeV/$c$.  The errors are statistical only.}
\label{bzero}
\begin{tabular}
{@{\hspace{0.5cm}}l|@{\hspace{0.5cm}}r@{{\hspace{0.15cm}}$\pm${\hspace{0.15cm}}}l@{\hspace{0.5cm}}@{\hspace{0.5cm}}r@{{\hspace{0.15cm}}$\pm${\hspace{0.15cm}}}l@{\hspace{0.5cm}}}


\hline \hline
$B$ candidate              & \multicolumn{2}{c}{$B^+$} \hspace{0.2cm}      & \multicolumn{2}{c}{$B^0$} \\
\hline \hline                        
On Resonance Data          & 6573 & 81            & 5564 & 75  \\   
\hline                                                                               
Scaled Off Resonance       & 258 & 48             & 218  & 45  \\   
Combinatorial Background   & 1394 & 38             & 765  & 28  \\
\hline                                                                              
Secondary                  & 680 & 26             & 1915 & 44  \\
\hline                                                                             
Background Subtracted      & 4241 & 105           & 2666 & 102 \\
\hline \hline
\end{tabular}
\end{table}

\section{The Electron Energy Spectrum and the Moments}
The electron energy spectrum is generated via Monte Carlo simulation of $B \rightarrow X_c e \nu$ decays with the EVTGEN event generator \cite{BelleMC}.   The spectrum from $B \rightarrow X_c e \nu$ is modelled using four components: 
$X_c = D$ (ISGW2~\cite{ref:8}), $D^*$ (HQET~\cite{ref:7}), higher resonance charm meson states $D^{**}$(ISGW2) 
and non-resonant $D^{(*)} \pi$ (Goity and Roberts~\cite{ref:9}). To account for the most recent theoretical and experimental results, we reweight the $D$ and $D^*$ components in $p_e^{*B}$ to the spectra generated with current (world) average form factors \cite{PDG}.  In addition, the branching fractions for the $D$ and $D^*$ components are corrected according to current (world) average values \cite{PDG}.
Electrons that come from the $b \to u$ transition are subtracted from
the unfolded electron energy spectrum, defined later. We model the
electron energy spectrum from $B \to X_u l \nu$ transitions using the
 De Fazio and Neubert prescription~\cite{dfn}. The $b$-quark motion parameters are derived in Ref. \cite{bsg}.  We scale according to the $B \to X_u l \nu$ branching fraction in Ref.~\cite{PDG}.

\begin{figure}[htb]
    \includegraphics[width=0.60\textwidth]{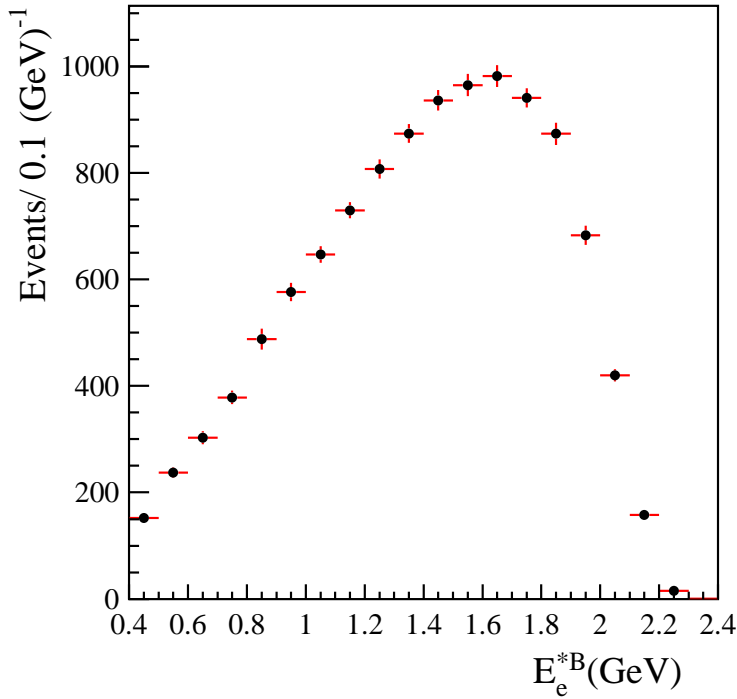}
  \caption{Unfolded electron energy distribution in the $B$ meson rest frame, combining contributions from $B^0$ and $B^+$ decays, and corrected for QED radiative effects. The errors shown are statistical.}
  \label{spectrum}
\end{figure}

To measure the first, second and third electron moments we need to determine the true electron energy spectrum in the $B$ meson rest frame, $E_{e}^{*B}$.  
The background subtracted momentum spectrum is distorted by various detector effects.
The true electron energy spectrum is extracted by performing an unfolding procedure based on 
the Singular Value Decomposition (SVD) algorithm~\cite{ref:13}. 
 The unfolded spectrum is corrected for QED radiative effects using the PHOTOS algorithm~\cite{PHOTOS}, as the OPE does not have an $\mathcal{O}(\alpha)$ QED correction.  The unfolded electron energy spectrum is shown in Figure \ref{spectrum}.

\begin{figure}[htb]
  \begin{tabular}{cc}
  \includegraphics[width=0.48\textwidth]{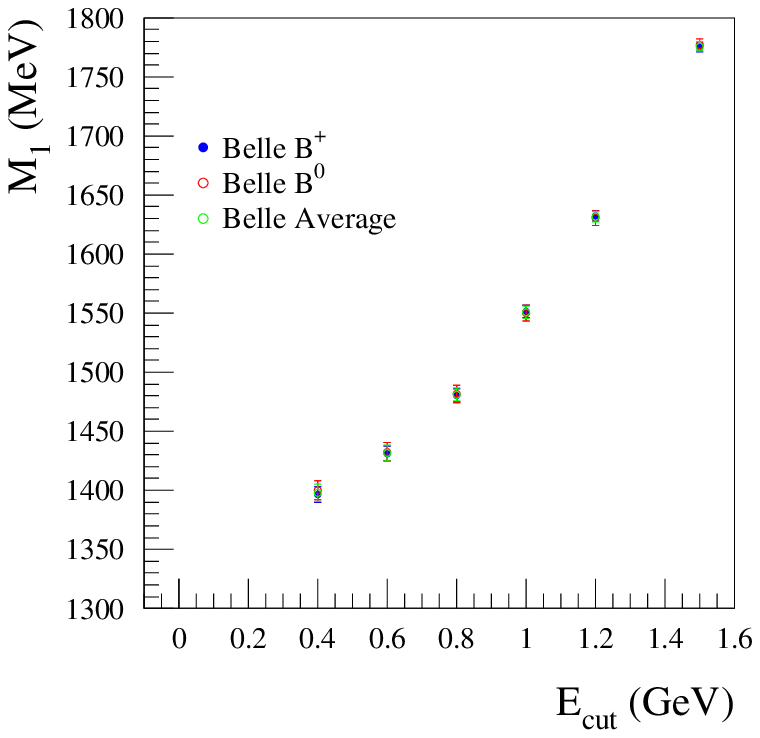}&
  \includegraphics[width=0.48\textwidth]{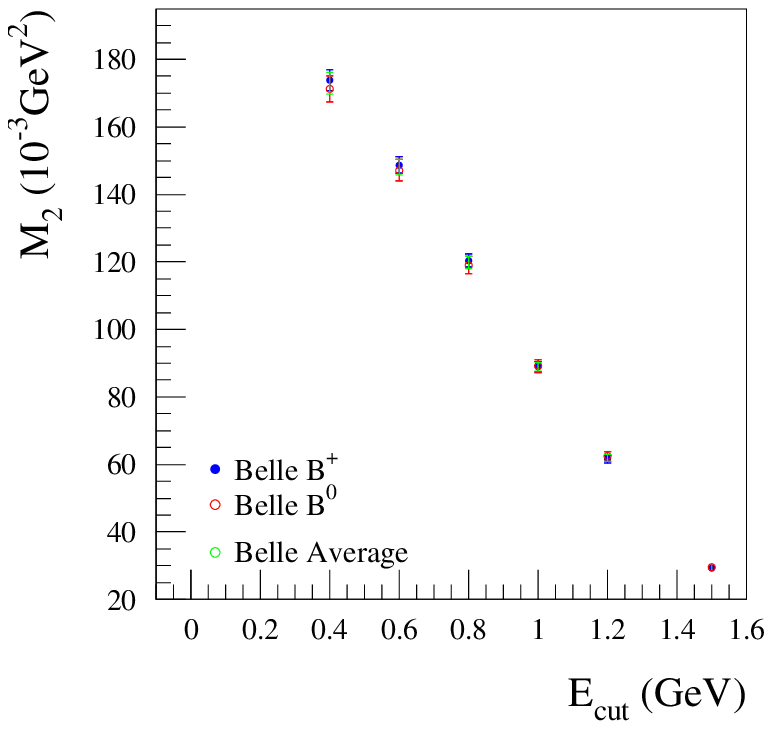}\\
  \includegraphics[width=0.48\textwidth]{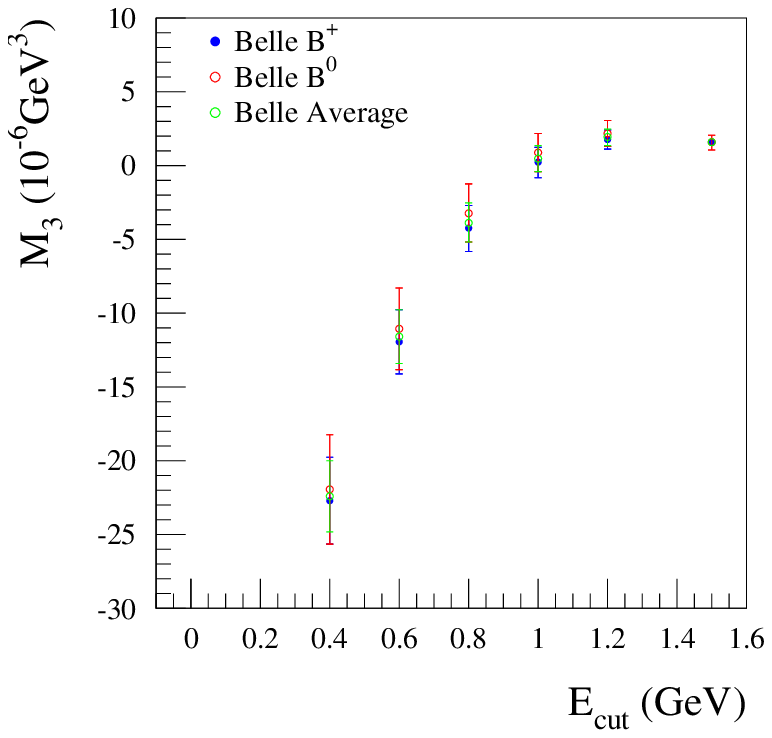}&\\
  \end{tabular}
\caption{First, second and third electron energy moments, $M_1$, $M_2$ and $M_3$, 
as a function of cutoff energy $E_{\mathrm{cut}}$ . The errors shown are statistical and
  systematic.}
\label{momentplots}
\end{figure}

We measure the central moments of the electron energy spectrum. The first  moment is defined to be
$M_1^I = \langle E_I \rangle$ and subsequent central moments are determined about the
first moment,
$M_n^I= \langle (E_I- \langle E_I\rangle )^n \rangle$,  where $I$ is the electron energy cutoff and $n=2,3$.
We measure the first three moments with six electron energy threshold cuts 
($I=E_{\mathrm{cut}}=$ 0.4, 0.6, 0.8, 1.0, 1.2 and 1.5 GeV) combining the spectra from $B^+$ and $B^0$ semileptonic decays.  
Table \ref{moments2} provides the final measurements of the moments.  Figure \ref{momentplots} shows the moments of the $B^0$ and $B^+$ subsamples as a function of $E_{\mathrm{cut}}$, as well as the $B^0$ and $B^+$ combined average.  

\section{Systematic Uncertainties}
The systematic uncertainties in the moments stem from event
selection, electron identification, background estimation and model
dependence. 

We determine the electron identification uncertainty from varying the electron identification constraints. In addition, we calculate a systematic uncertainty associated with the electron tracking and detection efficiencies.

Model dependence is estimated from the observed change
in the moments when the $B \to D e \nu$ and $B \to D^{*} e \nu$ decay
shape parameters are varied according to their uncertainties. 
The uncertainty in secondary ($B \rightarrow D \rightarrow e$) decays
is derived from switching the branching fraction corrections on and off.  
Similarly, the correction to the prompt  $B \to D^{(*)} e \nu$
branching fraction in the Monte Carlo, incurs such a systematic uncertainty.

The uncertainty due to mis-tagging in the $B^0$ and $B^+$ samples is
derived from the magnitude of the combinatorial
background. Uncertainty in the measured luminosity
adds to the uncertainty on continuum electron yield.  Additionally we account for the uncertainty associated to magnitude of the hadron fake contribution. 
The systematic due to the overall fit for the secondaries to the data is estimated by varying the lower $p_e^{*B}$ bound of the fit.

To estimate unfolding uncertainty we vary the effective rank parameter in the SVD algorithm. 
The uncertainty due to the $b \to u$ subtraction, which occurs after unfolding, is evaluated by varying the normalisation of the De Fazio$-$Neubert inclusive spectrum.

The total systematic error is obtained by adding each contribution in quadrature.  It is important to note that the systematic errors are limited by Monte Carlo statistics.
The contributions to the systematic error are summarised in Tables \ref{sys1}, \ref{sys2}, \ref{sys3} for the first, second and third moments respectively. 
\begin{table}[htb]
\caption{Measured moments, $M_1$, $M_2$, $M_3$ and the branching fraction for $B \rightarrow X_c e \nu$ in the $B$ meson rest frame for six cutoff energies $E_{\mathrm{cut}}$.  The first error is the statistical, and the second error is the systematic.  The moments are corrected for QED radiative effects.}
\label{moments2}
\begin{center}
\begin{tabular}
{@{\hspace{0.1cm}}c@{\hspace{0.2cm}}|@{\hspace{0.4cm}}r@{\hspace{0.4cm}}r@{\hspace{0.4cm}}r@{\hspace{0.4cm}}r@{\hspace{0.4cm}}r@{\hspace{0.4cm}}r@{\hspace{0.4cm}}r@{\hspace{0.4cm}}
}
\hline \hline
$E_{\mathrm{cut}}$[GeV]& $M_1$[MeV]\hspace{0.7cm} & $M_2$[$10^{-3}\rm GeV^2$]\hspace{0.2cm} & $M_3$[$10^{-6}\rm GeV^3$] \hspace{0.1cm} \\
\hline\hline
0.4 & 1397.7 $\pm$ 5.1 $\pm$ 5.4 & 172.8 $\pm$ 2.4 $\pm$ 2.2 &  -22.4 $\pm$ 2.3 $\pm$ 0.7  \\ 
0.6 & 1431.8 $\pm$ 4.8 $\pm$ 4.3 & 148.2 $\pm$ 1.9 $\pm$ 1.2 &  -11.6 $\pm$ 1.7 $\pm$ 0.6  \\
0.8 & 1481.0 $\pm$ 4.4 $\pm$ 3.4 & 119.9 $\pm$ 1.6 $\pm$ 0.9 &  -3.9  $\pm$ 1.2 $\pm$ 0.6  \\
1.0 & 1550.8 $\pm$ 4.0 $\pm$ 2.9 & 89.0  $\pm$ 1.2 $\pm$ 0.4 &   0.5  $\pm$ 0.8 $\pm$ 0.3  \\
1.2 & 1631.6 $\pm$ 3.6 $\pm$ 2.2 & 61.9  $\pm$ 0.9 $\pm$ 0.6 &   1.9  $\pm$ 0.5 $\pm$ 0.2  \\
1.5 & 1775.8 $\pm$ 3.0 $\pm$ 2.3 & 29.4  $\pm$ 0.6 $\pm$ 0.3 &   1.6  $\pm$ 0.2 $\pm$ 0.1  \\
\hline\hline
\end{tabular}
\end{center}
\end{table}

\begin{table}[htb]
\caption{Breakdown of the systematic errors for the first moment, $M_1$, for $B \rightarrow X_c e \nu$ in the $B$ meson rest frame for all 6 values of $E_{\mathrm{cut}}$}
\label{sys1}
\
\begin{tabular}
{l|@{\hspace{1.cm}}r@{\hspace{1.cm}}r@{\hspace{1.cm}}r@{\hspace{1.cm}}r@{\hspace{1.cm}}r@{\hspace{1.cm}}r}
\hline \hline
                               & $M_1$ & $M_1$ & $M_1$ & $M_1$ & $M_1$  & $M_1$ \\
                               & \footnotesize{[MeV]} & \footnotesize{[MeV]} & \footnotesize{[MeV]} & \footnotesize{[MeV]} & \footnotesize{[MeV]} & \footnotesize{[MeV]} \\
$E_{\mathrm{cut}}$[GeV]          &0.4&0.6&0.8&1.0&1.2&1.5\\
\hline
\hline
electron identification        &0.69&0.59&0.57&0.50&0.44&0.27\\
detection efficiency            &0.00&0.00&0.00&0.00&0.00&0.00\\
\hline
$B \to D^{(*)} e \nu$ form factors      &2.08&1.53&1.06&0.89&0.84&1.07\\
$B \to D^{(*)} e \nu$ Br                &4.06&3.69&3.04&2.61&2.00&1.50\\
$B \to D_{(s)}^{(*)} \to e$ Br           &0.21&0.18&0.13&0.08&0.05&0.01\\
\hline
continuum background           &0.35&0.35&0.30&0.19&0.08&0.04\\
combinatorial background       &0.01&0.06&0.06&0.04&0.03&0.06\\
hadron fakes                   &0.70&0.60&0.45&0.25&0.08&0.02\\
secondaries                    &1.99&0.79&0.67&0.49&0.01&1.30\\
\hline
unfolding                      &1.63&0.67&0.32&0.44&0.31&0.35\\
$b \to u$ subtraction          &0.17&0.19&0.21&0.21&0.21&0.21\\
\hline \hline
total systematics              &5.38&4.27&3.37&2.94&2.20&2.30\\
\hline\hline
\end{tabular}

\end{table}
\begin{table}[htb]
\caption{Breakdown of the systematic errors for the second moment, $M_2$, for $B \rightarrow X_c e \nu$ in the $B$ meson rest frame for all 6 values of $E_{\mathrm{cut}}$}
\label{sys2}

\begin{tabular}
{l|rrrrrr}
\hline \hline
                               & $M_2$ & $M_2$ & $M_2$ & $M_2$ & $M_2$  & $M_2$ \\
                               & \footnotesize{[$10^{-3}\rm GeV^2$]} & \footnotesize{[$10^{-3}\rm GeV^2$]} & \footnotesize{[$10^{-3}\rm GeV^2$]} & \footnotesize{[$10^{-3}\rm GeV^2$]} & \footnotesize{[$10^{-3}\rm GeV^2$]} & \footnotesize{[$10^{-3}\rm GeV^2$]} \\
$E_{\mathrm{cut}}$[GeV]          &0.4&0.6&0.8&1.0&1.2&1.5\\
\hline
\hline
electron identification        &0.31&0.26&0.2&0.15&0.11&0.05\\
detection efficiency            &0.00&0.00&0.00&0.00&0.00&0.00\\
\hline
$B \to D^{(*)} e \nu$ form factors  &0.81&0.43&0.32&0.29&0.27&0.16\\    
$B \to D^{(*)} e \nu$ Br      &0.18&0.07&0.22&0.22&0.22&0.10\\
$B \to D_{(s)}^{(*)} \to e$ Br &0.05&0.03&0.02&0.01&0.00&0.00\\
\hline
continuum background           &0.12&0.10&0.06&0.03&0.02&0.00\\
combinatorial background       &0.08&0.06&0.04&0.03&0.02&0.01\\
hadron fakes                   &0.17&0.12&0.07&0.03&0.01&0.00\\
secondaries                    &1.36&0.78&0.72&0.13&0.40&0.09\\
\hline
unfolding                      &1.43&0.78&0.25&0.16&0.20&0.21\\
$b \to u$ subtraction          &0.13&0.11&0.09&0.08&0.07&0.06\\
\hline \hline
total systematics              &2.18&1.19&0.88&0.42&0.57&0.31\\
\hline\hline
\end{tabular}
\end{table}
\begin{table}[htb]
\caption{Breakdown of the systematic errors for the third moment, $M_3$, for $B \rightarrow X_c e \nu$ in the $B$ meson rest frame for all 6 values of $E_{\mathrm{cut}}$}
\label{sys3}
\begin{tabular}
{l|rrrrrr}
\hline \hline
                               & $M_3$ & $M_3$ & $M_3$ & $M_3$ & $M_3$  & $M_3$ \\
                               & \footnotesize{[$10^{-6}\rm GeV^3$]} & \footnotesize{[$10^{-6}\rm GeV^3$]} & \footnotesize{[$10^{-6}\rm GeV^3$]} & \footnotesize{[$10^{-6}\rm GeV^3$]} & \footnotesize{[$10^{-6}\rm GeV^3$]} & \footnotesize{[$10^{-6}\rm GeV^3$]} \\
$E_{\mathrm{cut}}$[GeV]          &0.4&0.6&0.8&1.0&1.2&1.5\\
\hline
\hline
electron identification        &0.13&0.12&0.08&0.05&0.02&0.00\\
detection efficiency            &0.00&0.00&0.00&0.00&0.00&0.00\\
\hline
$B \to D^{(*)} e \nu$ form factors      &0.17&0.11&0.09&0.05&0.03&0.03\\
$B \to D^{(*)} e \nu$ Br      &0.49&0.42&0.33&0.21&0.12&0.05\\
$B \to D_{(s)}^{(*)} \to e$ Br &0.02&0.02&0.01&0.01&0.00&0.00\\
\hline
continuum background           &0.06&0.05&0.04&0.02&0.01&0.00\\
combinatorial background       &0.02&0.02&0.01&0.00&0.00&0.00\\
hadron fakes                   &0.07&0.05&0.04&0.02&0.01&0.00\\
secondaries                    &0.35&0.44&0.41&0.20&0.05&0.00\\
\hline
unfolding                      &0.24&0.06&0.05&0.08&0.07&0.04\\
$b \to u$ subtraction          &0.01&0.01&0.02&0.02&0.02&0.02\\
\hline \hline
total systematics              &0.72&0.63&0.55&0.31&0.15&0.07\\
\hline\hline
\end{tabular}
\end{table}

\begin{sidewaystable}[ptb]
\caption{Correlation matrix for 18 measured moments.  Correlation coefficients are statistical only.}
\label{cor}
\begin{center}

\begin{tabular}
{r|rrrrrr|rrrrrr|rrrrrr}
\hline\hline
          &$M_1^{0.4}$&$M_1^{0.6}$&$M_1^{0.8}$&$M_1^{1.0}$&$M_1^{1.2}$&$M_1^{1.5}$&$M_2^{0.4}$&$M_2^{0.6}$&$M_2^{0.8}$&$M_2^{1.0}$&$M_2^{1.2}$&$M_2^{1.5}$&$M_3^{0.4}$&$M_3^{0.6}$&$M_3^{0.8}$&$M_3^{1.0}$&$M_3^{1.2}$&$M_3^{1.5}$\\ 
\hline
$M_1^{0.4}$                   & 1.00 & 0.91 & 0.79 & 0.64 & 0.49 & 0.28 &-0.28 &-0.15 &-0.05 & 0.01 & 0.03 & 0.03 & 0.85 & 0.62 & 0.42 & 0.25 & 0.14 & 0.04 \\  
$M_1^{0.6}$   &                      & 1.00 & 0.87 & 0.71 & 0.54 & 0.31 &-0.18 &-0.19 &-0.06 & 0.01 & 0.04 & 0.04 & 0.77 & 0.87 & 0.59 & 0.35 & 0.19 & 0.06 \\
$M_1^{0.8}$   &&                            & 1.00 & 0.81 & 0.62 & 0.35 &-0.06 &-0.08 &-0.08 & 0.01 & 0.05 & 0.06 & 0.70 & 0.77 & 0.88 & 0.53 & 0.28 & 0.08 \\
$M_1^{1.0}$   &&&                                  & 1.00 & 0.77 & 0.43 & 0.01 & 0.01 & 0.02 & 0.01 & 0.07 & 0.08 & 0.56 & 0.62 & 0.71 & 0.89 & 0.46 & 0.14 \\
$M_1^{1.2}$   &&&&                                        & 1.00 & 0.56 & 0.05 & 0.06 & 0.07 & 0.08 & 0.12 & 0.12 & 0.43 & 0.46 & 0.53 & 0.66 & 0.86 & 0.25 \\
$M_1^{1.5}$   &&&&&                                              & 1.00 & 0.08 & 0.09 & 0.10 & 0.13 & 0.16 & 0.29 & 0.23 & 0.25 & 0.30 & 0.36 & 0.47 & 0.83 \\
\hline
$M_2^{0.4}$   &&&&&&                                                    & 1.00 & 0.80 & 0.60 & 0.41 & 0.27 & 0.11 &-0.48 &-0.22 &-0.05 & 0.07 & 0.05 & 0.06 \\
$M_2^{0.6}$   &&&&&&&                                                          & 1.00 & 0.75 & 0.52 & 0.34 & 0.14 &-0.25 &-0.31 &-0.06 & 0.04 & 0.06 & 0.08 \\
$M_2^{0.8}$   &&&&&&&&                                                                & 1.00 & 0.69 & 0.44 & 0.19 &-0.06 &-0.08 &-0.09 & 0.08 & 0.10 & 0.10 \\
$M_2^{1.0}$   &&&&&&&&&                                                                      & 1.00 & 0.64 & 0.27 & 0.03 & 0.05 & 0.06 & 0.11 & 0.16 & 0.15 \\
$M_2^{1.2}$   &&&&&&&&&&                                                                            & 1.00 & 0.42 & 0.07 & 0.10 & 0.13 & 0.19 & 0.26 & 0.23 \\
$M_2^{1.5}$   &&&&&&&&&&&                                                                                  & 1.00 & 0.02 & 0.04 & 0.06 & 0.09 & 0.16 & 0.55 \\
\hline
$M_3^{0.4}$   &&&&&&&&&&&&                                                                                        & 1.00 & 0.71 & 0.47 & 0.27 & 0.15 & 0.05 \\
$M_3^{0.6}$   &&&&&&&&&&&&&                                                                                              & 1.00 & 0.66 & 0.39 & 0.21 & 0.07 \\
$M_3^{0.8}$   &&&&&&&&&&&&&&                                                                                                    & 1.00 & 0.59 & 0.32 & 0.10 \\
$M_3^{1.0}$   &&&&&&&&&&&&&&&                                                                                                          & 1.00 & 0.54 & 0.16 \\
$M_3^{1.2}$   &&&&&&&&&&&&&&&&                                                                                                                & 1.00 & 0.30 \\
$M_3^{1.5}$   &&&&&&&&&&&&&&&&&                                                                                                                      & 1.00 \\

\hline\hline
\end{tabular}
\end{center}
\end{sidewaystable}

A set of statistical correlation coefficients for all measured moments with $E_{\rm cut}$ ranging from 0.4 GeV to 1.5 GeV are given in Table \ref{cor}.  

\section{Summary}
We report a measurement of the electron energy spectrum of the
inclusive decay $B \rightarrow X_c e \nu$ and its first, second and
third moments for threshold energies from 0.4 GeV to 1.5 GeV.  This set of moments, combined with the measurements of the semileptonic branching fraction and the moments of the hadronic  mass distribution, will be used for the determination of the HQE parameters and of $|V_{cb}|$.

\section{Acknowledgments}
We thank the KEKB group for the excellent operation of the accelerator, the KEK Cryogenics group for the efficient
operation of the solenoid, and the KEK computer group and the National Institute of Informatics for valuable computing
and Super-SINET network support. We acknowledge support from the Ministry of Education, Culture, Sports, Science, and
Technology of Japan and the Japan Society for the Promotion of Science; the Australian Research Council and the
Australian Department of Education, Science and Training; the National Science Foundation of China under contract No.~10175071; the Department of Science and Technology of India; the BK21 program of the Ministry of Education of
Korea and the CHEP SRC program of the Korea Science and Engineering Foundation; the Polish State Committee for
Scientific Research under contract No.~2P03B 01324; the Ministry of Science and Technology of the Russian
Federation; the Ministry of Education, Science and Sport of the Republic of Slovenia; the National Science Council and
the Ministry of Education of Taiwan; and the U.S. Department of Energy.

%





\begin{thebibliography}{99}
\bibitem{CC}
Throughout this paper, 
the inclusion of the charge conjugate mode decay is implied
unless otherwise stated.
\end{thebibliography}
\end{document}